\pdfoutput=1

\documentclass[11pt,twoside,a4paper,cmspaper,final,collab]{cms-tdr}
\def\svnVersion{281964:282295}\def\cmsCernNo{2015-075}\def\cmsCernDate{\today}\def\cmsMessage{Submitted to Physical Review Letters}

\begin{document}\cmsNoteHeader{HIG-14-042}

\hyphenation{had-ron-i-za-tion}
\hyphenation{cal-or-i-me-ter}
\hyphenation{de-vices}
\RCS$Revision: 282195 $
\RCS$HeadURL: svn+ssh://alverson@svn.cern.ch/reps/tdr2/papers/HIG-14-042/trunk/HIG-14-042.tex $
\RCS$Id: HIG-14-042.tex 282195 2015-03-25 14:18:15Z alverson $
\cmsNoteHeader{CMS-HIG-14-042}
\renewcommand{\cmsCollabName}{The ATLAS and CMS Collaborations}
\renewcommand{\cmsNUMBER}{HIG-14-042}
\renewcommand{\cmsPubBlock}{\begin{tabular}[t]{@{}r@{}l}&CMS \cmsSTYLE\
\cmsNUMBER\\&ATLAS-HIGG-2014-14\\\end{tabular}}
\renewcommand{\cmslogo}{\includegraphics[height=2.33cm]{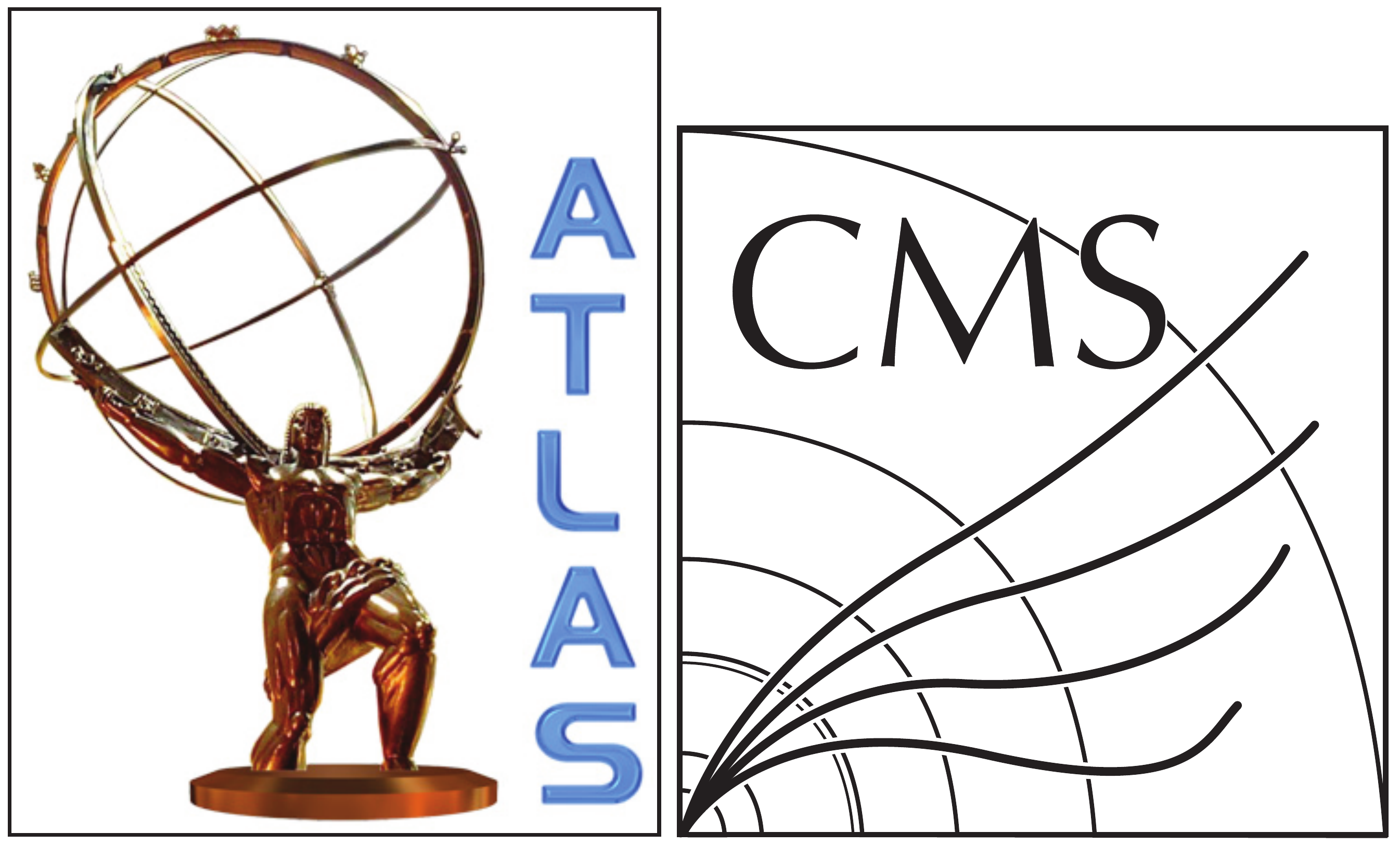}}
\renewcommand{\cmsTag}{CMS-\cmsNUMBER\\&ATLAS-HIGG-2014-14\\}
\renewcommand{\appMsg}{See appendices A and B for lists of collaboration members.}
\renewcommand{\cmsCopyright}{\copyright\,\the\year\ CERN for the benefit of the ATLAS and CMS Collaborations.}
\providecommand{\squeezetable}{\scriptsize}
\newlength\cmsFigWidth
\ifthenelse{\boolean{cms@external}}{\setlength\cmsFigWidth{0.75\textwidth}}{\setlength\cmsFigWidth{\textwidth}}

\title{Combined Measurement of the Higgs Boson Mass in \texorpdfstring{$pp$}{pp} Collisions at
\texorpdfstring{$\sqrt{s}=7$}{sqrt(s)=7} and 8 TeV with the ATLAS and CMS Experiments}

\author{The ATLAS and CMS Collaborations}

\date{\today}

\def\Hboson{\ensuremath{H}}
\def\antibar#1{\ensuremath{#1\bar{#1}}}
\def\tbar{\ensuremath{\bar{t}}}
\def\ttbar{\antibar{t}}
\def\bbbar{\antibar{b}}
\def\qqbar{\antibar{q}}
\def\vec#1{{\mbox{$\boldsymbol{#1}$}}}

\newcommand{\mgg}{\ensuremath{m_{\gamma\gamma}}\xspace}

\newcommand{\Amumu}{\ensuremath{h/H/A \rightarrow \mu^{+}\mu^{-}}}
\newcommand{\emu}{\ensuremath{e^{\pm}\mu^{\mp}}}
\newcommand{\mup}{\ensuremath{\mu^{+}}}
\providecommand{\mum}{}
\renewcommand{\mum}{\ensuremath{\mu^{-}}}
\newcommand{\fb}{\ensuremath{\,\mbox{fb}}}
\newcommand{\bbA}{\ensuremath{b\bar{b}A}}

\newcommand{\bb}{\ensuremath{b\bar{b}}}
\def\imb{\mbox{$\mu$b$^{-1}$}}

\def\pt{\ensuremath{p_{\mathrm{T}}}} 
\def\mt{\ensuremath{m_{\mathrm{T}}}} 
\def\mT{\ensuremath{m_{\mathrm{T}}}} 

\newcommand\ttH{\ensuremath{\ttbar\Hboson}}
\newcommand\bbH{\ensuremath{\bbbar\Hboson}}
\newcommand\ggH{\ensuremath{gg\mathrm{F}}}
\newcommand\VBFH{\ensuremath{\mathrm{VBF}}}
\newcommand\WH{\ensuremath{W\Hboson}}
\newcommand\ZH{\ensuremath{Z\Hboson}}
\newcommand\VH{\ensuremath{V\Hboson}}

\newcommand{\Zgjets}{$Z/\gamma^{*}$+jets}
\newcommand{\Zjets}{$Z$+jets}
\newcommand{\mll}{$m_{\ell\ell}$}
\newcommand{\dphill}{$\Delta\phi_{\ell\ell}$}

\def\hgg{\ensuremath{H \rightarrow \gamma\gamma}}
\def\hfourl{\ensuremath{H \rightarrow 4\ell}}
\newcommand{\hww}{\ensuremath{H \rightarrow WW^{(*)}}}
\def\hWWlnln{\ensuremath{H \rightarrow WW^{(*)} \rightarrow \ell\nu\ell\nu}}
\def\lnln{\ensuremath{\ell\nu\ell\nu}}
\def\hWWlnqq{\ensuremath{H \rightarrow WW \rightarrow \ell\nu qq}}
\newcommand{\lnqq}{\ensuremath{\ell\nu qq}}
\newcommand{\hzz}{\ensuremath{H \rightarrow ZZ^{(*)}}}
\newcommand{\hZZ}{\ensuremath{H \rightarrow ZZ}}
\def\hZZllnn{\ensuremath{H \rightarrow ZZ\rightarrow \ell\ell\nu\nu}}
\def\hZZllqq{\ensuremath{H \rightarrow ZZ\rightarrow \ell\ell q\bar{q}}}
\def\hZZllll{\ensuremath{H \rightarrow ZZ\rightarrow 4\ell}}
\newcommand{\htt}{\ensuremath{H \rightarrow \tau^+\tau^-}}
\newcommand{\httlh}{\ensuremath{H\rightarrow \tau\tau \rightarrow \ell \tau_{had } 3 \nu }}
\newcommand{\httllj}{\ensuremath{H\rightarrow \tau\tau \rightarrow \ell^+\ell^- + 4\nu }}
\newcommand{\hbb}{\ensuremath{H \rightarrow \bb}}
\newcommand{\llnn}{\ensuremath{\ell^+\ell^-\nu\bar{\nu}}}
\newcommand{\hllnn}{\ensuremath{H \rightarrow ZZ \rightarrow \llnn}}
\newcommand{\zzllnn}{\ensuremath{ZZ \rightarrow \llnn}}
\newcommand{\llll}{\ensuremath{\ell^+\ell^-\ell^+\ell^-}}
\newcommand{\hllll}{\ensuremath{H \rightarrow ZZ^{(*)} \rightarrow \llll}}
\newcommand{\llbb}{\ensuremath{\ell^+\ell^-b\bar{b}}}
\newcommand{\lvbb}{\ensuremath{\ell\nu b\bar{b}}}
\def\wbb{\ensuremath{Wb\bar{b}}}
\def\zbb{\ensuremath{Zb\bar{b}}}
\newcommand{\hllbb}{\ensuremath{H \rightarrow ZZ \rightarrow \llbb}}
\newcommand{\heenn}{\ensuremath{H \rightarrow ZZ \rightarrow e^+e^-\nu\bar{\nu}}}
\newcommand{\hmmnn}{\ensuremath{H \rightarrow ZZ \rightarrow \mu^+\mu^-\nu\bar{\nu}}}
\newcommand{\mH}{\ensuremath{m_{H}}}
\newcommand{\mHhat}{\ensuremath{\hat{m_{H}}}}
\newcommand{\mh}{\ensuremath{m_{H}}}
\newcommand{\ztoll}{\ensuremath{Z \rightarrow \ell^+\ell^-}}
\newcommand\tthbb{\ensuremath{t\overline{t}}\hbb}
\newcommand\htollbb{\hllbb}
\newcommand\htollnunu{\hllnn}
\newcommand\zztollnunu{\zzllnn}
\newcommand\htollll{\hllnn}
\newcommand{\htoeenunu}{\heenn}
\newcommand{\htomumununu}{\hmmnn}
\newcommand{\lumiuncertainty}{$\pm3.7$\%}

\newcommand{\ObsA}{{\color{black}141}}
\newcommand{\ObsB}{{\color{black}238}}
\newcommand{\ObsC}{{\color{black}252}}
\newcommand{\ObsD}{{\color{black}476}}

\newcommand{\ObsANine}{{\color{black}146}}
\newcommand{\ObsDNine}{{\color{black}443}}

\newcommand{\ExpA}{{\color{black}124}}
\newcommand{\ExpB}{{\color{black}520}}

\newcommand{\ExcessA}{{\color{red}XXX}}
\newcommand{\ExcessB}{{\color{black} 170}}

\newcommand{\pmin}{{\color{black}$\sim$0.001}}
\newcommand{\Zmax}{{\color{black} 3.1}}
\newcommand{\upcross}{{\color{black}6}}
\newcommand{\pglob}{{\color{black} 0.05}}
\newcommand{\Zglob}{{\color{black}1.6}}

\newcommand{\significance}{{\color{black}{$2-4 ~\sigma$}}}
\newcommand{\lhood}{\ensuremath{{\cal L}}}
\newcommand{\gaussprob}{\ensuremath{\textrm{Norm}(\delta_i|\bar{\delta}_i)}}
\newcommand{\nuisance}{\ensuremath{m_{\delta_i}}}
\newcommand{\ZZbkg}{\ensuremath{ZZ^{(*)}}}
\newcommand{\zz}{\ensuremath{ZZ}}
\newcommand{\ww}{\ensuremath{WW}}
\newcommand{\wz}{\ensuremath{WZ}}
\newcommand{\Wjets}{$W$+jets}
\newcommand{\SM}{Standard Model}

\newcommand\htollvv{$H\to ZZ\to\ell^+\ell^-\nu\bar{\nu}$\ }
\newcommand\htollqq{$H\to ZZ\to\ell^+\ell^- q \bar{q}$\ }
\newcommand\htoeevv{$H\to ZZ\to ee\nu\nu$\ }
\newcommand\htoeeqq{$H\to ZZ\to eeq \bar{q}$\ }
\newcommand\htouuvv{$H\to ZZ\to\mu\mu\nu\nu$\ }
\newcommand\htouuqq{$H\to ZZ\to\mu\mu q \bar{q}$\ }
\newcommand{\inpb}{pb$^{-1}$}
\newcommand{\infb}{fb$^{-1}$}

\newcommand{\expt}{\ensuremath{\mathrm{expt}}}
\newcommand{\dm}{\ensuremath{\Delta m}}
\newcommand{\dmexp}{\ensuremath{\Delta m^{\expt}}}
\newcommand{\dmzg}{\ensuremath{\Delta m_{Z\gamma}}}
\newcommand{\dmgz}{\ensuremath{\Delta m_{\gamma Z}}}
\newcommand{\mufg}{\ensuremath{\mu_{F\gamma}}}
\newcommand{\rvf}{\ensuremath{R_{VF}}}
\newcommand{\rzg}{\ensuremath{R_{Z\gamma}}}

\newcommand{\ATLASm}{\ensuremath{\mathrm{ATLAS}}}
\newcommand{\CMSm}{\ensuremath{\mathrm{CMS}}}

\newcommand{\rf}{\ensuremath{\mu_{\ggH+\ttH}^{\gamma\gamma}}}
\newcommand{\rfATLAS}{\ensuremath{\mu_{\ggH+\ttH}^{\gamma\gamma~\ATLASm}}}
\newcommand{\rfCMS}{\ensuremath{\mu_{\ggH+\ttH}^{\gamma\gamma~\CMSm}}}
\newcommand{\rfcms}{\ensuremath{\mu_{\ggH+\ttH}^{\gamma\gamma}}}
\newcommand{\rv}{\ensuremath{\mu_{\VBFH+\VH}^{\gamma\gamma}}}
\newcommand{\rvATLAS}{\ensuremath{\mu_{\VBFH+\VH}^{\gamma\gamma~\ATLASm}}}
\newcommand{\rvCMS}{\ensuremath{\mu_{\VBFH+\VH}^{\gamma\gamma~\CMSm}}}
\newcommand{\muzz}{\ensuremath{\mu^{4\ell}}}
\newcommand{\muzzATLAS}{\ensuremath{\mu^{4\ell~\ATLASm}}}
\newcommand{\muzzCMS}{\ensuremath{\mu^{4\ell~\CMSm}}}

\newcommand{\rfhat}{\ensuremath{\hat{\mu}_{\ggH+\ttH}^{\gamma\gamma}}}
\newcommand{\rvhat}{\ensuremath{\hat{\mu}_{\VBFH+\VH}^{\gamma\gamma}}}
\newcommand{\muzzhat}{\ensuremath{\hat{\mu}^{4\ell}}}

\newcommand{\lfg}{\ensuremath{\lambda_{F\gamma}^{\expt}}}
\newcommand{\lvf}{\ensuremath{\lambda_{VF}^{\expt}}}
\newcommand{\lzg}{\ensuremath{\lambda_{Z\gamma}^{\expt}}}

\newcommand{\lmu}{\ensuremath{\lambda^{\expt}}}
\newcommand{\lrf}{\ensuremath{\lambda_{F}^{\expt}}}
\newcommand{\lrv}{\ensuremath{\lambda_{V}^{\expt}}}
\newcommand{\lzz}{\ensuremath{\lambda_{4\ell}^{\expt}}}

\newcommand{\hwwlnln}{\ensuremath{H \rightarrow WW^{(*)} \rightarrow \ell\nu\ell\nu}}
\newcommand{\hwwlnqq}{\ensuremath{H \rightarrow WW \rightarrow \ell\nu qq}}

\providecommand{\gev}{GeV}

\def\ifb{\mbox{fb$^{-1}$}}
\def\ipb{\mbox{pb$^{-1}$}}
\def\inb{\mbox{nb$^{-1}$}}

\newcommand{\alphaS}{\ensuremath{\alpha_S}}
\newcommand{\mtop}{\ensuremath{m_t}}
\newcommand{\mbottom}{\ensuremath{m_b}}
\newcommand{\mcharm}{\ensuremath{m_c}}

\hypersetup{%
pdfauthor={ATLAS and CMS Collaborations},%
pdftitle={Combined Measurement of the Higgs Boson Mass in pp Collisions at
sqrt(s) = 7 and 8 TeV with the ATLAS and CMS Experiments},%
pdfsubject={ATLAS, CMS},%
pdfkeywords={ATLAS, CMS, physics, Higgs, mass}}

\abstract{
A measurement of the Higgs boson mass is presented
based on the combined data samples
of the ATLAS and CMS experiments at the CERN LHC
in the $H \rightarrow \gamma\gamma$ and $H \rightarrow ZZ\rightarrow 4\ell$ decay channels.
The results are obtained from a simultaneous fit to the
reconstructed invariant mass peaks in the two channels
and for the two experiments.
The measured masses from the individual channels and the two
experiments are found to be consistent among themselves.
The combined measured mass of the Higgs boson is
$m_{H} = 125.09\pm0.21\,\mathrm{(stat.)}\pm0.11\,\mathrm{(syst.)}~\mathrm{GeV}$.
}

\maketitle 

The study of the mechanism of electroweak symmetry
breaking is one of the principal goals of the CERN LHC program.
In the Standard Model (SM),
this symmetry breaking is achieved through the introduction of a
complex doublet scalar field,
leading to the prediction of the Higgs boson
$H$~\cite{Englert:1964et,Higgs:1964ia,Higgs:1964pj,Guralnik:1964eu,Higgs:1966ev,Kibble:1967sv},
whose mass \mH\ is, however, not predicted by the theory.
In 2012,
the ATLAS and CMS Collaborations at the LHC announced the discovery
of a particle with Higgs boson-like properties and a
mass of about 125~\gev~\cite{HiggsObservationATLAS, HiggsObservationCMS, CMSLong2013}.
The discovery was based primarily on mass peaks observed in the $\gamma\gamma$ and
$ZZ\rightarrow\ell^+\ell^-\ell^{\prime +}\ell^{\prime -}$
(denoted \hZZllll\ for simplicity) decay channels,
where one or both of the $Z$ bosons can be off-shell
and where $\ell$ and $\ell^\prime$ denote an electron or muon.
With \mH\ known,
all properties of the SM Higgs boson,
such as its production cross section and partial decay widths,
can be predicted.
Increasingly precise
measurements~\cite{atlas_coupling_paper,atlas_spin_paper,CMS_combination,Khachatryan:2014kca}
have established that all observed properties of the new particle,
including its spin, parity, and coupling strengths to SM particles
are consistent within the uncertainties with those expected for the SM Higgs boson.

The ATLAS and CMS Collaborations have independently measured \mH\ using
the samples of proton-proton collision data collected
in 2011 and 2012,
commonly referred to as LHC Run~1.
The analyzed samples correspond to approximately 5~\ifb\
of integrated luminosity at $\sqrt{s}=7$~TeV,
and 20~\ifb\ at $\sqrt{s}=8$~TeV,
for each experiment.
Combined results in the context of the separate experiments,
as well as those in the individual channels,
are presented in Refs.~\cite{atlas_mass_paper, CMS_Hgg, CMS_HZZ,CMS_combination}.

This Letter describes a combination of the Run~1 data from the two experiments,
leading to improved precision for~\mH.
Besides its intrinsic importance as a fundamental parameter,
improved knowledge of \mH\ yields more precise predictions for
the other Higgs boson properties.
Furthermore,
the combined mass measurement provides a first step towards combinations
of other quantities,
such as the couplings.
In the SM, \mH\ is related to the values of the masses of the
$W$ boson and top quark through loop-induced effects.
Taking into account other measured SM quantities,
the comparison of the measurements of the Higgs boson,
$W$ boson, and top quark masses can be used to directly
test the consistency of the SM~\cite{Baak:2014ora}
and thus to search for evidence of physics beyond the SM.

The combination is performed using only the \hgg\ and \hZZllll\ decay channels,
because these two channels offer the best mass resolution.
Interference between the Higgs boson signal and the continuum
background is expected to produce a downward shift
of the signal peak relative to the true value of~\mH.
The overall effect in the \hgg\ channel~\cite{Dixon:2003yb,Martin:2012xc,Dixon:2013haa}
is expected to be a few tens of MeV
for a Higgs boson with a width near the SM value,
which is small compared to the current precision.
The effect in the \hZZllll\ channel is expected to be much smaller~\cite{Kauer:2012hd}.
The effects of the interference on the mass spectra are neglected in this
Letter.

The ATLAS and CMS detectors~\cite{ATLAS,CMS} are
designed to precisely reconstruct charged leptons, photons,
hadronic jets, and the imbalance of momentum
transverse to the direction of the beams.
The two detectors are based on different technologies requiring
different reconstruction and calibration methods.
Consequently they are subject to different sources of systematic uncertainty.

The \hgg\ channel is characterized by a narrow resonant signal peak
containing several hundred events per experiment
above a large falling continuum background.
The overall signal-to-background ratio is a few percent.
Both experiments divide the
\hgg\ events into different categories depending
on the signal purity and mass resolution,
as a means to improve sensitivity.
While CMS uses the same analysis procedure for the measurement of
the Higgs boson mass and couplings~\cite{CMS_Hgg},
ATLAS implements separate analyses
for the couplings~\cite{atlas_hgg_coupling}
and for the mass~\cite{atlas_mass_paper};
the latter analysis classifies events
in a manner that reduces the expected systematic uncertainties in~\mH.

The \hZZllll\ channel yields only a few tens of signal events per experiment,
but has very little background,
resulting in a signal-to-background ratio larger than~1.
The events are analyzed separately depending on
the flavor of the lepton pairs.
To extract \mH, ATLAS employs a two-dimensional (2D) fit to the
distribution of the four-lepton mass and a kinematic discriminant introduced
to reject the main background,
which arises from
$ZZ$ continuum production.
The CMS procedure is based on a three-dimensional fit,
utilizing the four-lepton mass distribution,
a kinematic discriminant,
and the estimated event-by-event uncertainty in the four-lepton mass.
Both analyses are optimized for the mass measurement and neither attempts to
distinguish between different Higgs boson production mechanisms.

There are only minor differences in the parameterizations
used for the present combination
compared to those used for the combination of
the two channels by the individual experiments.
These differences have almost no effect on the results.

The measurement of \mh,
along with its uncertainty,
is based on the maximization of profile-likelihood ratios $\Lambda(\vec\alpha)$
in the asymptotic regime~\cite{LHC-HCG,Cowan:2010st}:
\begin{linenomath}\begin{equation}
  \Lambda(\vec\alpha) = {\frac{L\big(\vec\alpha\,,\,\hat{\hat{\vec\theta}}(\vec\alpha)\big)}
                              {L(\hat{\vec\alpha},\hat{\vec\theta})},
\label{eq:LH}}
\end{equation}\end{linenomath}
where $L$ represents the likelihood function,
$\vec\alpha$ the parameters of interest,
and $\vec\theta$ the nuisance parameters.
There are three types of nuisance parameters:
those corresponding to systematic uncertainties,
the fitted parameters of the background models,
and any unconstrained signal model parameters not relevant to the particular hypothesis under test.
Systematic uncertainties
are discussed below.
The other two types of nuisance parameters are incorporated
into the statistical uncertainty.
The
$\vec\theta$ terms are profiled,
i.e., for each possible value of a parameter of interest
(e.g., \mH),
all nuisance parameters are refitted to maximize~$L$.
The $\hat{\vec\alpha}$ and $\hat{\vec\theta}$ terms denote the unconditional
maximum likelihood estimates of the best-fit values for the parameters,
while $\hat{\hat{\vec\theta}}(\vec\alpha)$ is the conditional
maximum likelihood estimate for given parameter values~$\vec\alpha$.

The likelihood functions $L$ are constructed using signal and background
probability density functions (PDFs) that depend on the
discriminating variables:
for the \hgg\ channel, the diphoton mass and,
for the \hZZllll\ channel,
the four-lepton mass (for CMS, also its uncertainty)
and the kinematic discriminant.
The signal PDFs are derived from samples of Monte Carlo (MC) simulated events.
For the \hZZllll\ channel,
the background PDFs are determined using a combination of
simulation and data control regions.
For the \hgg\ channel,
the background PDFs are obtained directly from the fit to the data.
The profile-likelihood fits to the data are performed
as a function of \mH\ and the signal-strength scale factors defined below.
The fitting framework is implemented independently by ATLAS and CMS,
using the {\sc RooFit}~\cite{Verkerke:2003ir}, {\sc RooStats}~\cite{Moneta:2010pm},
and {\sc HistFactory}~\cite{Cranmer:2012sba} data modeling and handling packages.

Despite the current agreement between the measured
Higgs boson properties and the SM predictions,
it is pertinent to perform a mass measurement that is as independent
as possible of SM assumptions.
For this purpose, three
signal-strength scale factors are introduced
and profiled
in the fit, thus reducing the dependence of the results
on assumptions about the Higgs boson couplings and
about the variation of the production cross section times
branching fraction with the mass.
The signal strengths are defined as
$\mu=(\sigma_{\expt}\times\mathrm{BF}_{\expt})/(\sigma_\mathrm{SM}\times\mathrm{BF}_\mathrm{SM})$,
representing the ratio of the
cross section times branching fraction
in the experiment
to the corresponding SM expectation
for the different production and decay modes.
Two factors,
$\rf$ and $\rv$,
are used to scale the signal strength in the \hgg\ channel.
The production processes involving Higgs boson couplings to fermions,
namely gluon fusion ($\ggH$) and associated production
with a top quark-antiquark pair ($\ttH$),
are scaled with the $\rf$ factor.
The production processes involving couplings to vector bosons,
namely vector boson fusion ($\VBFH$) and associated production with a vector boson ($\VH$),
are scaled with the $\rv$ factor.
The third factor, $\muzz$,
is used to scale the signal strength in the \hZZllll\ channel.
Only a single signal-strength parameter is used
for \hZZllll\ events because
the \mH\ measurement in this case is found to exhibit
almost no sensitivity to the different production mechanisms.

The procedure based on the two scale factors $\rf$ and $\rv$
for the \hgg\ channel
was previously employed by CMS~\cite{CMS_Hgg}
but not by ATLAS.
Instead, ATLAS relied on a single \hgg\
signal-strength scale factor.
The additional degree-of-freedom introduced by ATLAS for
the present study results in a shift of about 40~MeV in
the ATLAS \hgg\ result,
leading to a shift of 20~MeV in the ATLAS combined mass measurement.

The individual signal strengths $\rf$, $\rv$, and $\muzz$
are assumed to be the same for ATLAS and CMS,
and are profiled in the combined fit for \mH.
The corresponding profile-likelihood ratio is
\ifthenelse{\boolean{cms@external}}{
\begin{widetext}
\begin{linenomath}\begin{equation}\label{eq:teststatMh}
  \Lambda(m_H) = \frac{L\big(m_H\,,\,\hat{\hat{\mu}}^{\gamma\gamma}_{\ggH+\ttH}(m_H)\,,\hat{\hat{\mu}}^{\gamma\gamma}_{\VBFH+\VH}(m_H)\,,\,\hat{\hat{\mu}}^{4\ell}(m_H)\,,\,\hat{\hat{\vec\theta}}(m_H)\big)} {L(\hat{m}_H,\hat{\mu}^{\gamma\gamma}_{\ggH+\ttH}\,,\hat{\mu}^{\gamma\gamma}_{\VBFH+\VH},\hat{\mu}^{4\ell},\hat{\vec\theta})} \;.
\end{equation}\end{linenomath}
\end{widetext}
}{
\begin{linenomath}\begin{equation}\label{eq:teststatMh}
  \Lambda(m_H) = \frac{L\big(m_H\,,\,\hat{\hat{\mu}}^{\gamma\gamma}_{\ggH+\ttH}(m_H)\,,\hat{\hat{\mu}}^{\gamma\gamma}_{\VBFH+\VH}(m_H)\,,\,\hat{\hat{\mu}}^{4\ell}(m_H)\,,\,\hat{\hat{\vec\theta}}(m_H)\big)} {L(\hat{m}_H,\hat{\mu}^{\gamma\gamma}_{\ggH+\ttH}\,,\hat{\mu}^{\gamma\gamma}_{\VBFH+\VH},\hat{\mu}^{4\ell},\hat{\vec\theta})} \;.
\end{equation}\end{linenomath}
}
Slightly more complex fit models are used,
as described below,
to perform additional compatibility tests between the
different decay channels and between the results from ATLAS and~CMS.

Combining the ATLAS and CMS data for the \hgg\ and \hZZllll\ channels
according to the above procedure,
the mass of the Higgs boson is determined to be
\begin{linenomath}\begin{equation}
\label{eq:massresult}
  \begin{split}
    \mH&=125.09\pm0.24~\mathrm{\gev}\\
    &=125.09\pm0.21\,\mathrm{(stat.)}\pm0.11\,\mathrm{(syst.)}~\mathrm{\gev},
  \end{split}
\end{equation}\end{linenomath}
where the total uncertainty is obtained from the width
of a negative log-likelihood ratio
scan with all parameters profiled.
The statistical uncertainty is determined by fixing all nuisance parameters
to their best-fit values,
except for the three signal-strength scale factors
and the \hgg\ background function parameters,
which are profiled.
The systematic uncertainty is determined by subtracting
in quadrature the statistical uncertainty from the total uncertainty.
Equation~(\ref{eq:massresult}) shows that
the uncertainties in the \mH\ measurement
are dominated by the statistical term,
even when the Run~1 data sets of ATLAS and CMS are combined.
Figure~\ref{figure_LHC_combined_obs} shows the
negative log-likelihood ratio scans as a function of {\mH},
with all nuisance parameters profiled (solid curves),
and with the nuisance parameters fixed
to their best-fit values (dashed curves).

The signal strengths at the measured value of \mH\ are found to be
$\rf=1.15^{+0.28}_{-0.25}$,\break $\rv=1.17^{+0.58}_{-0.53}$,
and $\muzz=1.40^{+0.30}_{-0.25}$.
The combined overall signal strength $\mu$
(with $\rf=\rv=\muzz\equiv\mu$) is $\mu=1.24^{+0.18}_{-0.16}$.
The results reported here for the signal strengths are not expected
to have the same sensitivity, nor exactly the same values, as those that
would be extracted from a combined analysis optimized
for the coupling measurements.

\begin{figure}[th]
  \centering
  \includegraphics[width=0.9\columnwidth]{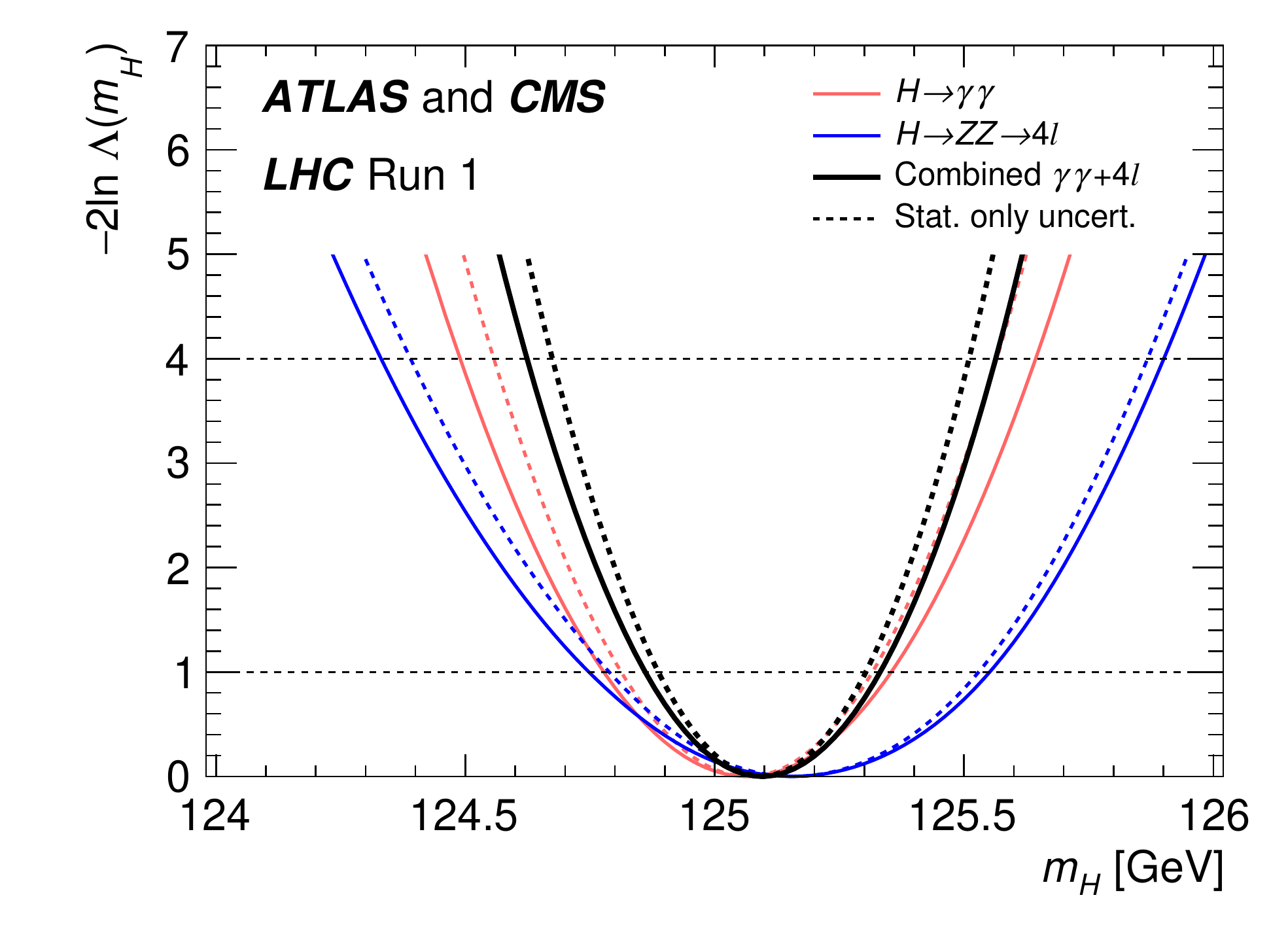}
   \caption{Scans of twice the negative log-likelihood ratio
$-2\ln\Lambda(\mH)$ as functions of the Higgs boson mass \mH\ for the
ATLAS and CMS combination of the \hgg\ (red), \hZZllll\ (blue),
and combined (black) channels.
The dashed curves show the results accounting for statistical uncertainties
only, with all nuisance parameters associated with systematic
uncertainties fixed to their best-fit values.
The 1 and 2 standard deviation limits are indicated by the
intersections of the horizontal lines at 1 and 4, respectively,
with the log-likelihood scan curves.
}
\label{figure_LHC_combined_obs}
\end{figure}

The combined ATLAS and CMS results for \mH\
in the separate \hgg\ and \hZZllll\ channels are
\begin{linenomath}\begin{equation}
  \begin{split}
    \mH^{\gamma\gamma}&=125.07\pm0.29~\mathrm{\gev}\\
    &=125.07\pm0.25\,\mathrm{(stat.)}\pm0.14\,\mathrm{(syst.)}~\mathrm{\gev}
  \end{split}
\end{equation}\end{linenomath}
and
\begin{linenomath}\begin{equation}
  \begin{split}
    \mH^{4\ell}&=125.15\pm0.40~\mathrm{\gev}\\
    &=125.15\pm0.37\,\mathrm{(stat.)}\pm0.15\,\mathrm{(syst.)}~\mathrm{\gev}.
  \end{split}
\end{equation}\end{linenomath}
The corresponding likelihood
ratio scans are
shown in Fig.~\ref{figure_LHC_combined_obs}.

A summary of the results from the individual
analyses and their combination is presented in
Fig.~\ref{figure_LHC_combined_obs_summary}.

\begin{figure*}[th]
  \centering
    \includegraphics[width=0.9\textwidth]{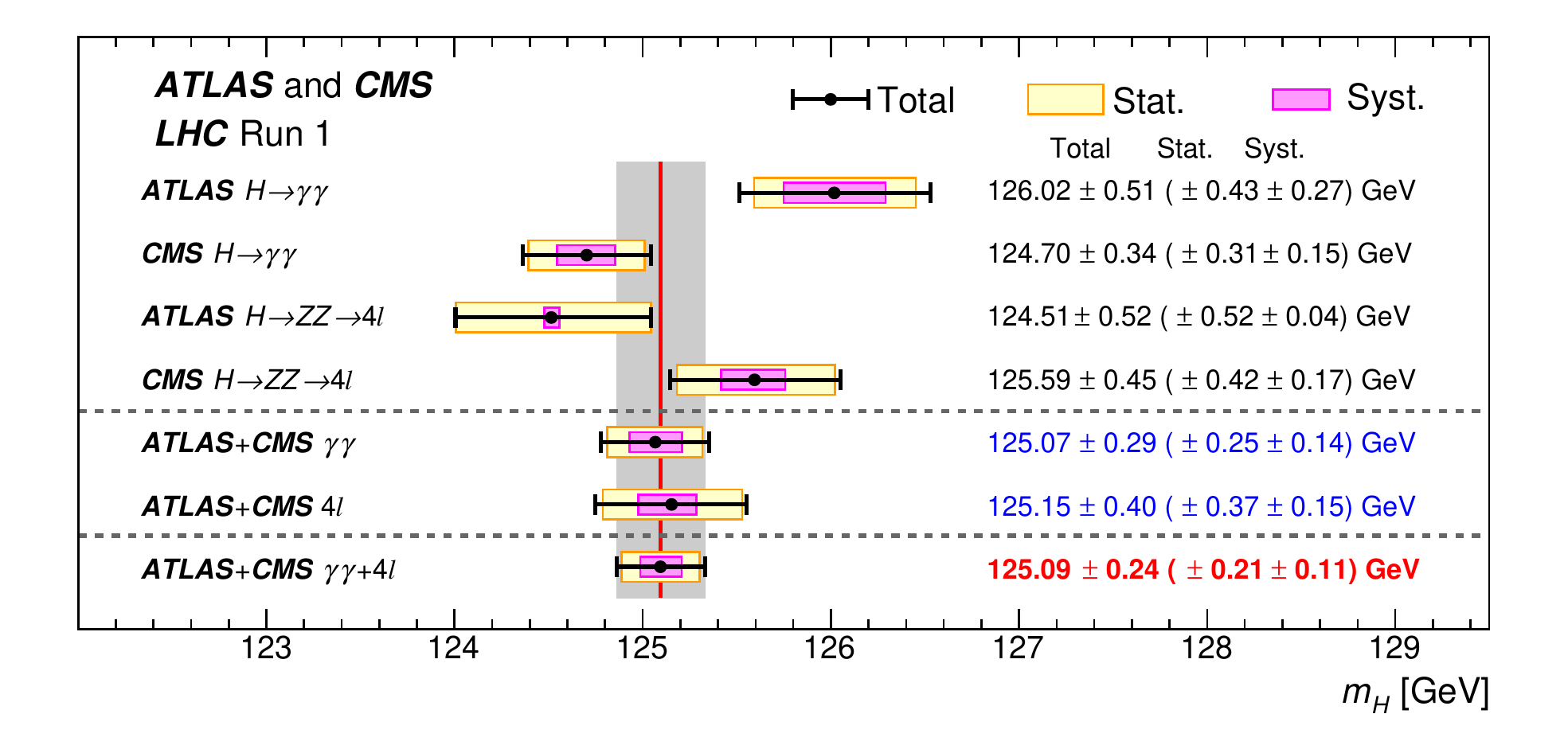}
\caption{Summary of Higgs boson mass measurements
from the individual analyses of ATLAS and CMS
and from the combined analysis presented here.
The systematic (narrower, magenta-shaded bands),
statistical (wider, yellow-shaded bands),
and total (black error bars) uncertainties are indicated.
The (red) vertical line and corresponding (gray) shaded column indicate the central value and the total uncertainty of the combined measurement, respectively.
}
\label{figure_LHC_combined_obs_summary}
\end{figure*}

The observed uncertainties in the combined measurement
can be compared with
expectations. The latter
are evaluated by generating
two Asimov data sets~\cite{Cowan:2010st},
where an Asimov data set is a representative event sample
that provides both the median expectation for an experimental result and
its expected statistical variation,
in the asymptotic approximation,
without the need for an extensive MC-based calculation.
The first Asimov data set is a ``prefit'' sample,
generated using $\mH=125.0$~GeV and the SM predictions for the couplings,
with all nuisance parameters fixed to their nominal values.
The second Asimov data set is a ``postfit'' sample,
in which \mh, the three signal strengths \rf, \rv, and \muzz, and all nuisance
parameters are fixed to their best-fit estimates
from the data.
The expected uncertainties for the combined mass are
\begin{linenomath}\begin{equation}
\delta {\mH}_\text{prefit}=\pm{0.24}~\rm{GeV}=\pm{0.22}\,\mathrm{(stat.)} \pm {0.10}\,\mathrm{(syst.)}~\rm{GeV}
\end{equation}\end{linenomath}
for the prefit case and
\begin{linenomath}\begin{equation}
\delta{\mH}_\text{postfit}=\pm{0.22}~\rm{GeV}=\pm{0.19}\,\mathrm{(stat.)}\,\pm{0.10} \mathrm{(syst.)}~\rm{GeV}
\end{equation}\end{linenomath}
for the postfit case, which are both very similar to the
observed uncertainties reported in Eq.~(\ref{eq:massresult}).

Constraining all signal yields to their SM predictions results
in an {\mH} value that is about 70~MeV larger than the nominal
result with a comparable uncertainty.
The increase in the central value reflects the combined effect of the
higher-than-expected \hZZllll\ measured signal strength
and the increase of the $H\rightarrow ZZ$ branching fraction with {\mH}.
Thus, the
fit assuming SM couplings forces the mass to a higher value
in order to accommodate the value $\mu=1$ expected in the SM.

Since the discovery,
both experiments have improved their understanding of the electron,
photon, and muon
measurements~\cite{Aad:2014nim,Aad:2014rra,Chatrchyan:2012xi,CMS_HZZ,Khachatryan:2015hwa,Khachatryan:2015iwa},
leading to a significant reduction of the systematic uncertainties
in the mass measurement.
Nevertheless, the treatment and understanding of systematic uncertainties
is an important aspect of the individual measurements and their combination.
The combined analysis incorporates approximately 300 nuisance parameters.
Among these, approximately 100 are fitted parameters describing the shapes and
normalizations of the background models in the \hgg\ channel,
including a number of discrete parameters
that allow the functional form in each of the CMS \hgg\ analysis
categories to be changed~\cite{Dauncey:2014xga}.
Of the remaining almost 200 nuisance parameters,
most correspond to experimental or theoretical systematic uncertainties.

Based on the results from the individual experiments,
the dominant systematic uncertainties for the combined \mH\ result
are expected to be those associated
with the energy or momentum scale and its resolution:
for the photons in the \hgg\ channel and for the
electrons and muons in the \hZZllll\ channel~\cite{atlas_mass_paper, CMS_Hgg, CMS_HZZ}.
These uncertainties are assumed to be uncorrelated between
the two experiments since they are related to the specific characteristics
of the
detectors as well as to the calibration procedures,
which are fully independent
except for negligible effects due to the
use of the common $Z$ boson mass~\cite{Z-Pole}
to specify
the absolute energy and momentum scales.
Other experimental systematic uncertainties~\cite{atlas_mass_paper, CMS_Hgg, CMS_HZZ}
are similarly assumed to be uncorrelated between the two experiments.
Uncertainties in the theoretical predictions
and in the measured integrated luminosities
are treated as fully and partially correlated,
respectively.

\begin{table*}
  \centering
\topcaption{\small Systematic uncertainties $\delta {\mH}$ (see text)
associated with the indicated effects
for each of the four input channels,
and the corresponding contributions of ATLAS and CMS
to the systematic uncertainties of the combined result.
``ECAL'' refers to the electromagnetic calorimeters.
The numbers in parentheses indicate expected values obtained from the prefit
Asimov data set discussed in the text.
The uncertainties for the combined result are related to the values
of the individual channels through the
relative weight of the individual channel in the combination,
which is proportional to the inverse of the
respective uncertainty squared.
The top section of the table divides the sources of systematic
uncertainty into three classes,
which are discussed in the text.
The bottom section of the table shows the total systematic uncertainties
estimated by adding the individual contributions in quadrature,
the total systematic uncertainties evaluated using
the nominal method discussed in the text,
the statistical uncertainties, the total uncertainties, and the analysis weights,
illustrative of the relative weight of each channel
in the combined \mH\ measurement.
}
  \resizebox{\textwidth}{!}{
  \begin{tabular}{lcccccc}
\hline
                                                       & \multicolumn{2}{c}{Uncertainty in ATLAS}& \multicolumn{2}{c}{Uncertainty in CMS}  & \multicolumn{2}{c}{Uncertainty in}      \\
                                                       & \multicolumn{2}{c}{results [GeV]:}      & \multicolumn{2}{c}{results [GeV]:}      & \multicolumn{2}{c}{combined result [GeV]:} \\
                                                       & \multicolumn{2}{c}{observed (expected)} & \multicolumn{2}{c}{observed (expected)} & \multicolumn{2}{c}{observed (expected)} \\
                                                       & \hgg               & \hZZllll           & \hgg               & \hZZllll           & ATLAS & CMS                             \\\hline
\hline
Scale uncertainties: & & & & & & \\
    ATLAS ECAL non-linearity /& & & & & & \\
    CMS~~~\,photon non-linearity                       &    0.14    (0.16)  &         --         &    0.10    (0.13)  &         --         &    0.02    (0.04)  &    0.05    (0.06)  \\
    Material in front of ECAL                          &    0.15    (0.13)  &         --         &    0.07    (0.07)  &         --         &    0.03    (0.03)  &    0.04    (0.03)  \\
    ECAL longitudinal response                         &    0.12    (0.13)  &         --         &    0.02    (0.01)  &         --         &    0.02    (0.03)  &    0.01    (0.01)  \\
    ECAL lateral shower shape                          &    0.09    (0.08)  &         --         &    0.06    (0.06)  &         --         &    0.02    (0.02)  &    0.03    (0.03)  \\
    Photon energy resolution                           &    0.03    (0.01)  &         --         &    0.01 ($<$0.01)  &         --         &    0.02 ($<$0.01)  & $<$0.01 ($<$0.01)  \\
    ATLAS \hgg\ vertex \& conversion   &    0.05    (0.05)  &         --         &         --         &         --         &    0.01    (0.01)  &         --         \\
    \quad reconstruction & & & & & & \\
    $Z\to ee$ calibration                              &    0.05    (0.04)  &    0.03    (0.02)  &    0.05    (0.05)  &         --         &    0.02    (0.01)  &    0.02    (0.02)  \\
    CMS electron energy scale \& resolution            &         --         &         --         &         --         &    0.12    (0.09)  &         --         &    0.03    (0.02)  \\
    Muon momentum scale \& resolution                  &         --         &    0.03    (0.04)  &         --         &    0.11    (0.10)  & $<$0.01    (0.01)  &    0.05    (0.02)  \\
\hline
Other uncertainties: & & & & & & \\
    ATLAS \hgg\ background                    &    0.04    (0.03)  &         --         &         --         &         --         &    0.01    (0.01)  &         --         \\
    \quad modeling& & & & & & \\
    Integrated luminosity                              &    0.01 ($<$0.01)  & $<$0.01 ($<$0.01)  &    0.01 ($<$0.01)  & $<$0.01 ($<$0.01)  &   \multicolumn{2}{c}{   0.01 ($<$0.01)} \\
    Additional experimental systematic    &    0.03 ($<$0.01)  & $<$0.01 ($<$0.01)  &    0.02 ($<$0.01)  &    0.01 ($<$0.01)  &    0.01 ($<$0.01)  &    0.01 ($<$0.01)  \\
uncertainties& & & & & & \\
\hline
    Theory uncertainties                               & $<$0.01 ($<$0.01)  & $<$0.01 ($<$0.01)  &    0.02 ($<$0.01)  & $<$0.01 ($<$0.01)  &   \multicolumn{2}{c}{   0.01 ($<$0.01)} \\
\hline\hline
    Systematic uncertainty (sum in              &    0.27    (0.27)  &    0.04    (0.04)  &    0.15    (0.17)  &    0.16    (0.13)  &   \multicolumn{2}{c}{   0.11    (0.10)} \\
    \quad quadrature)& & & & & & \\
    Systematic uncertainty (nominal)                   &    0.27    (0.27)  &    0.04    (0.05)  &    0.15    (0.17)  &    0.17    (0.14)  &   \multicolumn{2}{c}{   0.11    (0.10)} \\
    Statistical uncertainty                            &    0.43    (0.45)  &    0.52    (0.66)  &    0.31    (0.32)  &    0.42    (0.57)  &   \multicolumn{2}{c}{   0.21    (0.22)} \\
    Total uncertainty                                  &    0.51    (0.52)  &    0.52    (0.66)  &    0.34    (0.36)  &    0.45    (0.59)  &   \multicolumn{2}{c}{   0.24    (0.24)} \\
    Analysis weights                                   &    19\%    (22\%)  &    18\%    (14\%)  &    40\%    (46\%)  &    23\%    (17\%)  &   \multicolumn{2}{c}{        --       } \\

 \end{tabular}
  }
  \label{tab:syst_grouping_channel}
\end{table*}

To evaluate the relative importance of
the different sources of systematic uncertainty,
the nuisance parameters are grouped according to their correspondence
to three broad classes of systematic uncertainty:
\begin{itemize}
\item uncertainties in the energy or momentum scale and
resolution for photons, electrons, and muons (``scale''),
\item theoretical uncertainties, e.g.,
uncertainties in the Higgs boson cross section and branching fractions,
and in the normalization of SM background processes
(``theory''),
\item other experimental uncertainties (``other'').
\end{itemize}
First, the total uncertainty is obtained from the full profile-likelihood scan,
as explained above.
Next, parameters associated with the ``scale'' terms are
fixed and a new scan is performed.
Then, in addition to the scale terms,
the parameters associated with the ``theory'' terms are fixed and a scan performed.
Finally,
in addition, the ``other'' parameters are fixed and a scan performed.
Thus the fits are performed iteratively,
with the different classes of nuisance parameters
cumulatively
held fixed to their best-fit values.
The uncertainties associated with the different
classes of nuisance parameters are defined by the
difference in quadrature between the uncertainties resulting from consecutive scans.
The statistical uncertainty is determined from the final scan,
with all
nuisance parameters associated with systematic terms held fixed,
as explained above.
The result is
\ifthenelse{\boolean{cms@external}}{
\begin{linenomath}\begin{multline}
\mH=125.09\pm 0.21\, (\rm{stat.})\pm 0.11\, (\rm{scale})\pm0.02\, (\rm{other})\\
 \pm0.01\, (\rm{theory}) ~\rm{GeV},
 \end{multline}\end{linenomath}
}{
\begin{linenomath}\begin{equation}
\mH=125.09\pm 0.21\, (\rm{stat.})\pm 0.11\, (\rm{scale})\pm0.02\, (\rm{other}) \pm0.01\, (\rm{theory}) ~\rm{GeV},
\end{equation}\end{linenomath}
}
\noindent from which it is seen that the systematic uncertainty is
indeed dominated by the energy and momentum scale terms.

The relative importance of the various sources of systematic uncertainty
is further investigated by
dividing the nuisance parameters into yet-finer groups,
with each group associated with a specific underlying effect,
and evaluating the impact of each group on the overall mass uncertainty.
The matching of nuisance parameters to an effect
is not strictly rigorous because nuisance parameters in the
two experiments do not always represent exactly the same
effect and in some cases multiple effects are related
to the same nuisance parameter.
Nevertheless the relative impact of the different effects
can be explored.
A few experiment-specific groups of nuisance parameters are defined.
For example, ATLAS includes a group of nuisance parameters to account for the inaccuracy of
the background modeling for the \hgg\ channel. To model this background,
ATLAS uses specific analytic functions in each category~\cite{atlas_mass_paper}
while CMS simultaneously considers different background parameterizations~\cite{Dauncey:2014xga}.
The systematic uncertainty in \mH\ related to the background modeling in CMS
is estimated to be negligible~\cite{CMS_Hgg}.

The impact of groups of nuisance parameters is evaluated starting
from the contribution of each individual nuisance parameter to the total uncertainty.
This contribution is defined as the mass shift $\delta {\mH}$ observed
when re-evaluating the profile-likelihood ratio after
fixing the nuisance parameter in question to its best-fit
value increased or decreased by 1~standard deviation ($\sigma$)
in its distribution.
For a nuisance parameter whose PDF is a Gaussian distribution,
this shift corresponds to the contribution of that particular
nuisance parameter to the final uncertainty.
The impact of a group of nuisance parameters
is estimated by summing in quadrature the contributions from the
individual parameters.

The impacts $\delta {\mH}$
due to each of the considered effects
are listed in Table~\ref{tab:syst_grouping_channel}.
The results are reported for the four individual channels,
both for the data and (in parentheses) the prefit Asimov data set.
The row labeled ``Systematic uncertainty (sum in quadrature)''
shows the total sums in quadrature of the individual terms in the table.
The row labeled ``Systematic uncertainty (nominal)"
shows the corresponding total systematic uncertainties derived using the subtraction in quadrature method discussed in connection with Eq.~(\ref{eq:massresult}).
The two methods to evaluate the total systematic uncertainty are seen to agree within 10~MeV,
which is comparable with the precision of the estimates.
The two rightmost columns of Table~\ref{tab:syst_grouping_channel} list the
contribution of each group of nuisance parameters to the uncertainties
in the combined mass measurement,
for ATLAS and CMS separately.

The statistical and total uncertainties are summarized
in the bottom section of Table~\ref{tab:syst_grouping_channel}.
Since the weight of a channel in the final combination is
determined by the inverse of the squared uncertainty,
the approximate relative weights for the combined result
are 19\% (\hgg) and 18\% (\hZZllll) for ATLAS,
and 40\% (\hgg) and 23\% (\hZZllll) for CMS.
These weights are reported in the last row of Table~\ref{tab:syst_grouping_channel},
along with the expected values.

Figure~\ref{figure_LHC_combined_NP_ranking1} presents
the impact of each group of nuisance parameters on the
total systematic uncertainty in the mass measurement of ATLAS, CMS,
and the combination.
For the individual ATLAS and CMS measurements,
the results in Fig.~\ref{figure_LHC_combined_NP_ranking1}
are approximately equivalent to the sum in quadrature of the respective
$\delta {\mH}$ terms in
Table~\ref{tab:syst_grouping_channel} multiplied by their analysis weights,
after normalizing these weights to correspond to either ATLAS only or CMS only.
The ATLAS and CMS combined results in
Fig.~\ref{figure_LHC_combined_NP_ranking1}
are the sum in quadrature of the combined results in
Table~\ref{tab:syst_grouping_channel}.

The results in Table~\ref{tab:syst_grouping_channel}
and Fig.~\ref{figure_LHC_combined_NP_ranking1}
establish that the largest systematic effects for the mass
uncertainty are those related to the determination of the energy scale of the photons,
followed by those associated with the determination of the electron
and muon momentum scales.
Since the CMS \hgg\ channel has the largest weight in the combination,
its impact on the systematic uncertainty of the combined result is largest.

\begin{figure*}[tbh]
  \centering
    \includegraphics[width=\cmsFigWidth]{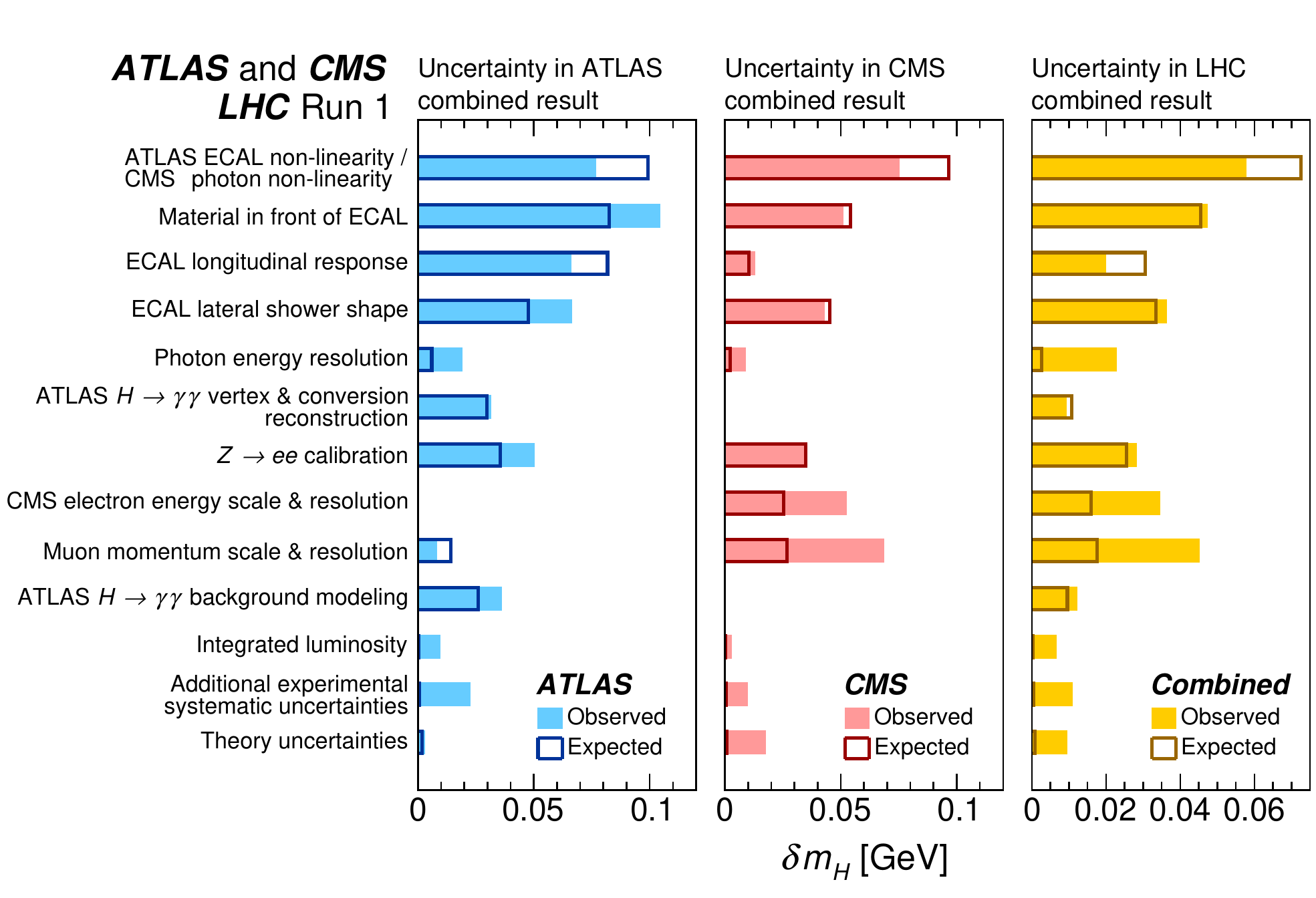}
    \caption{
The impacts $\delta\mH$ (see text)
of the nuisance parameter groups in Table~\ref{tab:syst_grouping_channel}
on the ATLAS (left), CMS (center), and combined (right) mass measurement uncertainty.
The observed (expected) results are shown by the solid (empty) bars.
}
\label{figure_LHC_combined_NP_ranking1}
\end{figure*}

The mutual compatibility of the \mH\ results from the four individual channels
is tested using a likelihood ratio with four masses in the numerator
and a common mass in the denominator,
and thus three degrees of freedom.
The three signal strengths are profiled in both the numerator
and denominator as in Eq.~(\ref{eq:LH}).
The resulting compatibility, defined as the asymptotic $p$-value of the fit,
is 10\%.
Allowing the ATLAS and CMS signal strengths to vary independently
yields a compatibility of 7\%.
This latter fit results in an \mH\ value that is 40~MeV
larger than the nominal result.

The compatibility of the combined ATLAS and CMS mass measurement in the \hgg\ channel
with the combined measurement in the \hZZllll\ channel is evaluated using the
variable
$\dmgz\equiv \mH^{\gamma\gamma}-\mH^{4\ell}$
as the parameter of interest,
with all other parameters, including~\mH, profiled.
Similarly, the compatibility of the ATLAS combined mass measurement
in the two channels with the CMS combined measurement in the two channels
is evaluated using the variable $\dmexp\equiv \mH^{\ATLASm}-\mH^{\CMSm}$.
The observed results,
$\dmgz=-0.1\pm0.5~\mathrm{\gev}$ and $\dmexp=0.4\pm0.5~\mathrm{\gev}$,
are both consistent with zero within 1$\,\sigma$.
The difference between the mass values in the two experiments
is $\dmexp_{\gamma\gamma}=1.3\pm0.6~\mathrm{\gev}$ (2.1$\,\sigma$)
for the \hgg\ channel and
$\dmexp_{4\ell}=-0.9\pm0.7~\mathrm{\gev}$ (1.3$\,\sigma$)
for the \hZZllll\ channel.
The combined results exhibit a greater degree of compatibility
than the results from the individual decay channels
because the {\dmexp} value has opposite signs in the two channels.

The compatibility of the signal strengths from ATLAS and CMS
is evaluated through the ratios
$\lambda^{\expt}=\mu^{\ATLASm}/\mu^{\CMSm}$,
$\lrf=\rfATLAS/\rfCMS$, and $\lzz=\muzzATLAS/\muzzCMS$.
For this purpose,
each ratio is individually taken to be the parameter of interest,
with all other nuisance parameters profiled,
including the remaining two ratios for the first two tests.
We find
$\lambda^{\expt}=1.21^{+0.30}_{-0.24}$, $\lrf=1.3^{+0.8}_{-0.5}$, and $\lzz=1.3^{+0.5}_{-0.4}$,
all of which are consistent with unity within 1$\,\sigma$.
The ratio $\lrv=\rvATLAS/\rvCMS$
is omitted
because the ATLAS mass measurement in the \hgg\ channel
is not sensitive to $\rv/\rf$.

The correlation between the signal strength and the measured mass is explored
with 2D likelihood scans as functions of $\mu$ and $\mH$.
The three signal strengths are assumed to be the same:
$\rf=\rv=\muzz\equiv\mu$,
and thus the ratios of the production cross sections times branching fractions are constrained to the SM predictions.
Assuming that the negative log-likelihood ratio
$-2\ln\Lambda(\mu,\mH)$ is distributed as a $\chi^2$ variable
with two degrees of freedom, the 68\% confidence level (CL)
confidence regions are shown
in Fig.~\ref{figure_LHC_combined_obs_2d} for each individual measurement,
as well as for the combined result.

\begin{figure}[th]
  \centering
    \includegraphics[width=0.9\columnwidth]{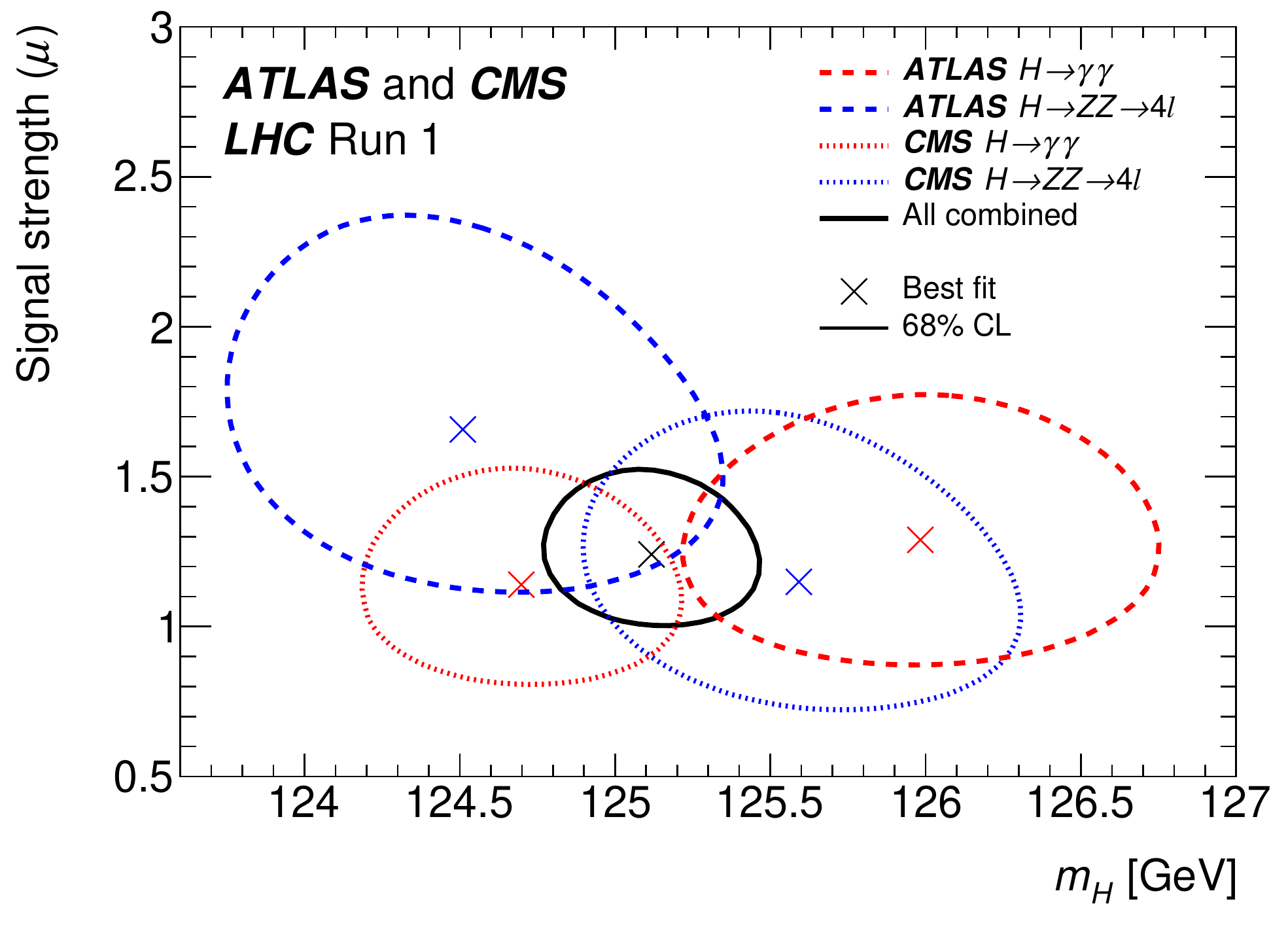}
\caption{Summary of likelihood scans in the 2D plane of
signal strength $\mu$ versus Higgs boson mass \mH\
for the ATLAS and CMS experiments.
The 68\% CL confidence regions of the individual measurements are shown by the dashed curves and of the overall combination by the solid curve.
The markers indicate the respective best-fit values.
}
\label{figure_LHC_combined_obs_2d}
\end{figure}

In summary, a combined measurement of the Higgs boson mass is performed in the \hgg\ and \hZZllll\ channels using the LHC Run~1 data sets of the ATLAS and CMS experiments, with minimal reliance on the assumption that the Higgs boson behaves as predicted by the SM.

The result is
\begin{linenomath}\begin{equation}
  \label{final_result}
  \begin{split}
    \mH &=125.09\pm0.24~\mathrm{\gev}\\
&=125.09\pm0.21\,\mathrm{(stat.)}\pm0.11\,\mathrm{(syst.)}~\mathrm{\gev},
  \end{split}
\end{equation}\end{linenomath}
where the total uncertainty is dominated by the statistical term,
with the systematic uncertainty dominated by effects related to
the photon, electron, and muon energy or momentum scales and resolutions.
Compatibility tests are performed to
ascertain
whether the measurements are consistent
with each other,
both between the different decay channels and between
the two experiments.
All tests on the combined results indicate consistency
of the different measurements within 1$\,\sigma$,
while the four Higgs boson mass measurements
in the two channels of the two experiments agree within 2$\,\sigma$.
The combined measurement of the Higgs boson mass
improves upon the results from the individual experiments
and is the most precise measurement to date of this fundamental parameter of
the newly discovered particle.

\section*{Acknowledgments}
We thank CERN for the very successful operation of the LHC, as well as
the support staff from our institutions without whom ATLAS and CMS
could not be operated efficiently.

We acknowledge the support of ANPCyT (Argentina); YerPhI (Armenia);
ARC (Australia); BMWFW and FWF (Austria); ANAS (Azerbaijan); SSTC
(Belarus); FNRS and FWO (Belgium); CNPq, CAPES, FAPERJ, and FAPESP
(Brazil); MES (Bulgaria); NSERC, NRC, and CFI (Canada); CERN; CONICYT
(Chile); CAS, MoST, and NSFC (China); COLCIENCIAS (Colombia); MSES and
CSF (Croatia); RPF (Cyprus); MSMT CR, MPO CR and VSC CR (Czech
Republic); DNRF, DNSRC and Lundbeck Foundation (Denmark); MoER, ERC
IUT and ERDF (Estonia);\break EPLANET, ERC and NSRF (European Union);
Academy of Finland, MEC, and HIP (Finland); CEA, CNRS/IN2P3 (France);
GNSF (Georgia); BMBF , DFG, HGF, MPG, and AvH Foundation (Germany);
GSRT and NSRF (Greece); RGC (Hong Kong SAR, China); OTKA and NIH
(Hungary); DAE and DST (India); IPM (Iran); SFI (Ireland); ISF,
MINERVA, GIF, I-CORE and Benoziyo Center (Israel); INFN (Italy); MEXT
and JSPS (Japan); JINR; MSIP and NRF (Republic of Korea); LAS
(Lithuania); MOE and UM (Malaysia); CINVESTAV, CONACYT, SEP, and
UASLP-FAI (Mexico); CNRST (Morocco); FOM and NWO (Netherlands); MBIE
(New Zealand); BRF and RCN (Norway); PAEC (Pakistan); MNiSW, MSHE,
NCN, and NSC (Poland); GRICES and FCT (Portugal); MNE/IFA (Romania);
MES of Russia, MON, RosAtom, RAS, and RFBR  (Russian Federation); MSTD
and MESTD (Serbia); MSSR (Slovakia); ARRS and MIZ\v{S} (Slovenia);
DST/NRF (South Africa); MINECO, SEIDI and CPAN (Spain); SRC and
Wallenberg Foundation (Sweden); ETH Board, ETH Zurich, PSI, SER, SNSF, 
UniZH, and Cantons of Bern, Gen\`eve and Zurich (Switzerland);
NSC (Taipei); MST (Taiwan); ThEPCenter, IPST, STAR and NSTDA
(Thailand); TUBITAK and TAEK (Turkey); NASU and SFFR (Ukraine); STFC
and the Royal Society and Leverhulme Trust (United Kingdom); DOE and
NSF (United States of America).

In addition, we gratefully acknowledge  the crucial computing support
from all  WLCG partners, in particular from CERN and the Tier-1 and
Tier-2 facilities worldwide.

\bibliography{auto_generated}   

\cleardoublepage \appendix\section{The ATLAS Collaboration \label{app:collab}}\begin{sloppypar}\hyphenpenalty=5000\widowpenalty=500\clubpenalty=5000G.~Aad$^{\rm 85}$,
B.~Abbott$^{\rm 113}$,
J.~Abdallah$^{\rm 151}$,
O.~Abdinov$^{\rm 11}$,
R.~Aben$^{\rm 107}$,
M.~Abolins$^{\rm 90}$,
O.S.~AbouZeid$^{\rm 158}$,
H.~Abramowicz$^{\rm 153}$,
H.~Abreu$^{\rm 152}$,
R.~Abreu$^{\rm 30}$
% ...
\bigskip
\\
% ...
$^{11}$ Institute of Physics, Azerbaijan Academy of Sciences, Baku, Azerbaijan\\
% ...
$^{30}$ CERN, Geneva, Switzerland\\
% ...
$^{85}$ CPPM, Aix-Marseille Universit{\'e} and CNRS/IN2P3, Marseille, France\\
% ...
$^{90}$ Department of Physics and Astronomy, Michigan State University, East Lansing MI, United States of America\\
% ...
$^{107}$ Nikhef National Institute for Subatomic Physics and University of Amsterdam, Amsterdam, Netherlands\\
% ...
$^{113}$ Homer L. Dodge Department of Physics and Astronomy, University of Oklahoma, Norman OK, United States of America\\
% ...
$^{151}$ Institute of Physics, Academia Sinica, Taipei, Taiwan\\
$^{152}$ Department of Physics, Technion: Israel Institute of Technology, Haifa, Israel\\
$^{153}$ Raymond and Beverly Sackler School of Physics and Astronomy, Tel Aviv University, Tel Aviv, Israel\\
% ...
$^{158}$ Department of Physics, University of Toronto, Toronto ON, Canada\\
% ...
$^{*}$ Deceased\clearpage\section{The CMS Collaboration}V.~Khachatryan$^{1}$,
A.M.~Sirunyan$^{1}$,
A.~Tumasyan$^{1}$,
W.~Adam$^{2}$,
E.~Asilar$^{2}$,
T.~Bergauer$^{2}$,
J.~Brandstetter$^{2}$,
E.~Brondolin$^{2}$,
M.~Dragicevic$^{2}$,
J.~Er\"{o}$^{2}$
% ...
\\[2ex]
$^{1}$~Yerevan Physics Institute, Yerevan, Armenia\\
$^{2}$~Institut f\"{u}r Hochenergiephysik der OeAW, Wien, Austria\\
% ...
\hrulefill\\[1ex]
\textit{\footnotesize
a~Deceased\\ 
b~Also~at~Vienna University of Technology, Vienna, Austria\\
c~Also~at~CERN, European Organization for Nuclear Research, Geneva, Switzerland\\
d~Also~at~State Key Laboratory of Nuclear Physics and Technology, Peking University, Beijing, China\\
e~Also~at~Institut Pluridisciplinaire Hubert Curien, Universit\'{e} de Strasbourg, Universit\'{e} de Haute Alsace Mulhouse, CNRS/IN2P3, Strasbourg, France\\
f~Also~at~National Institute of Chemical Physics and Biophysics, Tallinn, Estonia\\
g~Also~at~Skobeltsyn Institute of Nuclear Physics, Lomonosov Moscow State University, Moscow, Russia\\
h~Also~at~Universidade Estadual de Campinas, Campinas, Brazil\\
i~Also~at~Centre National de la Recherche Scientifique (CNRS) - IN2P3, Paris, France\\
j~Also~at~Laboratoire Leprince-Ringuet, Ecole Polytechnique, IN2P3-CNRS, Palaiseau, France\\
k~Also~at~Joint Institute for Nuclear Research, Dubna, Russia\\
m~Now~at~British University in Egypt, Cairo, Egypt\\
n~Now~at~Helwan University, Cairo, Egypt\\
o~Also~at~Suez University, Suez, Egypt\\
p~Also~at~Cairo University, Cairo, Egypt\\
q~Now~at~Fayoum University, El-Fayoum, Egypt\\
s~Now~at~Ain Shams University, Cairo, Egypt\\
u~Also~at~Universit\'{e} de Haute Alsace, Mulhouse, France\\
v~Also~at~Brandenburg University of Technology, Cottbus, Germany\\
w~Also~at~Institute of Nuclear Research ATOMKI, Debrecen, Hungary\\
x~Also~at~E\"{o}tv\"{o}s Lor\'{a}nd University, Budapest, Hungary\\
y~Also~at~University of Debrecen, Debrecen, Hungary\\
z~Also~at~Wigner Research Centre for Physics, Budapest, Hungary\\
aa~Also~at~University of Visva-Bharati, Santiniketan, India\\
bb~Now~at~King Abdulaziz University, Jeddah, Saudi Arabia\\
cc~Also~at~University of Ruhuna, Matara, Sri Lanka\\
dd~Also~at~Isfahan University of Technology, Isfahan, Iran\\
ee~Also~at~University of Tehran, Department of Engineering Science, Tehran, Iran\\
ff~Also~at~Plasma Physics Research Center, Science and Research Branch, Islamic Azad University, Tehran, Iran\\
gg~Also~at~Universit\`{a} degli Studi di Siena, Siena, Italy\\
hh~Also~at~Purdue University, West Lafayette, USA\\
ii~Also~at~International Islamic University of Malaysia, Kuala Lumpur, Malaysia\\
jj~Also~at~CONSEJO NATIONAL DE CIENCIA Y TECNOLOGIA, MEXICO, Mexico\\
kk~Also~at~Institute for Nuclear Research, Moscow, Russia\\
ll~Also~at~Institute of High Energy Physics and Informatization, Tbilisi State University, Tbilisi, Georgia\\
mm~Also~at~St. Petersburg State Polytechnical University, St. Petersburg, Russia\\
nn~Also~at~National Research Nuclear University 'Moscow Engineering Physics Institute' (MEPhI), Moscow, Russia\\
oo~Also~at~California Institute of Technology, Pasadena, USA\\
pp~Also~at~Faculty of Physics, University of Belgrade, Belgrade, Serbia\\
qq~Also~at~Facolt\`{a} Ingegneria, Universit\`{a} di Roma, Roma, Italy\\
rr~Also~at~National Technical University of Athens, Athens, Greece\\
ss~Also~at~Scuola Normale e Sezione dell'INFN, Pisa, Italy\\
tt~Also~at~University of Athens, Athens, Greece\\
uu~Also~at~Warsaw University of Technology, Institute of Electronic Systems, Warsaw, Poland\\
vv~Also~at~Institute for Theoretical and Experimental Physics, Moscow, Russia\\
ww~Also~at~Albert Einstein Center for Fundamental Physics, Bern, Switzerland\\
xx~Also~at~Adiyaman University, Adiyaman, Turkey\\
yy~Also~at~Mersin University, Mersin, Turkey\\
zz~Also~at~Cag University, Mersin, Turkey\\
aaa~Also~at~Piri Reis University, Istanbul, Turkey\\
bbb~Also~at~Gaziosmanpasa University, Tokat, Turkey\\
ccc~Also~at~Ozyegin University, Istanbul, Turkey\\
ddd~Also~at~Izmir Institute of Technology, Izmir, Turkey\\
eee~Also~at~Mimar Sinan University, Istanbul, Istanbul, Turkey\\
fff~Also~at~Marmara University, Istanbul, Turkey\\
ggg~Also~at~Kafkas University, Kars, Turkey\\
hhh~Also~at~Yildiz Technical University, Istanbul, Turkey\\
iii~Also~at~Hacettepe University, Ankara, Turkey\\
jjj~Also~at~Rutherford Appleton Laboratory, Didcot, United Kingdom\\
kkk~Also~at~School of Physics and Astronomy, University of Southampton, Southampton, United Kingdom\\
lll~Also~at~Instituto de Astrof\'{i}sica de Canarias, La Laguna, Spain\\
mmm~Also~at~Utah Valley University, Orem, USA\\
nnn~Also~at~University of Belgrade, Faculty of Physics and Vinca Institute of Nuclear Sciences, Belgrade, Serbia\\
ooo~Also~at~Argonne National Laboratory, Argonne, USA\\
ppp~Also~at~Erzincan University, Erzincan, Turkey\\
qqq~Also~at~Texas A\&M University at Qatar, Doha, Qatar\\
rrr~Also~at~Kyungpook National University, Daegu, Korea\\
}\end{sloppypar}
\end{document}